\documentclass[preprint,aps,showpacs,10pt]{revtex4-1}
\usepackage{hyperref}
\usepackage{amsmath}
\usepackage{amsfonts}
\usepackage{amssymb}
\usepackage[dvipdf]{graphicx}
\def\gsim{\lower.7ex\hbox{$\;\stackrel{\textstyle>}{\sim}\;$}}
\def\lsim{\lower.7ex\hbox{$\;\stackrel{\textstyle<}{\sim}\;$}}
\usepackage{graphicx}
\usepackage{dcolumn}
\usepackage{bm}
\usepackage{xcolor}
\usepackage{hyperref}
\usepackage[mathlines]{lineno}
\usepackage{subfigure}
\def\gsim{\lower.7ex\hbox{$\;\stackrel{\textstyle>}{\sim}\;$}}
\def\lsim{\lower.7ex\hbox{$\;\stackrel{\textstyle<}{\sim}\;$}}
\usepackage{color}
\usepackage[utf8]{inputenc}
\usepackage{soul}
\usepackage{ulem}
\definecolor{red}{rgb}{1,0,0}
\usepackage{epsfig,bbm}
\newcommand{\ba}{\begin{array}}
\newcommand{\ea}{\end{array}}
\newcommand{\be}{\begin{equation}}
\newcommand{\ee}{\end{equation}}
\newcommand{\bea}{\begin{eqnarray}}
\newcommand{\eea}{\end{eqnarray}}

\usepackage{mathtools}
\usepackage{multirow} 

\begin{document}
\title{CP breaking in $S(3)$ flavoured Higgs model}
\author{E. Barradas-Guevara}
 \email{barradas@fcfm.buap.mx}
\affiliation{Fac. de Cs. F\'{\i}sico Matem\'aticas, Benem\'erita Universidad Aut\'onoma de Puebla, Apdo. Postal 1152, Puebla, Pue.  72000, M\'exico.}%
\author{O. F\'elix-Beltr\'an}
\altaffiliation[Also at ]{olga\_flix@ece.buap.mx}
\affiliation{Fac. de Cs. de la Electr\'onica, Benem\'erita Universidad Aut\'onoma de Puebla, Apdo. Postal 542, Puebla, Pue. 72000, M\'exico.}
\author{E. Rodr\'{\i}guez-J\'auregui}
 \email{ezequiel.rodriguez@correo.fisica.uson.mx}
\affiliation{Departamento de F\'{\i}sica, Universidad de Sonora, Apdo. Postal 1626, Hermosillo, Son.  83000, M\'exico.}%

\date{\today}

\begin{abstract}
We analyze the Higgs sector of the minimal $S(3)$-invariant extension of the Standard Model including CP violation arising from the spontaneous electroweak symmetry breaking. This extended Higgs sector includes three SU(2) doublets Higgs fields with complex vev's providing an interesting scenario to analyze the Higgs masses spectrum, trilinear Higgs self-couplings and CP violation. We present how the spontaneous electroweak symmetry breaking coming from three $S(3)$ Higgs fields gives an interesting scenario with nine physical Higgs and three Goldstone bosons when spontaneous CP violation arises from the Higgs field $S(3)$ singlet $H_S$. Furthermore, a numerical analysis of the Higgs masses and trilinear Higgs self-couplings is presented. Particularly, we find a physical solution for the scenario in which spontaneous CPB is provided by $H_S$. In this scheme, the scalar Higgs $H^0_1$ is identified, whose mass is 125 GeV and $\lambda_{H_{1}^0 H_{1}^0 H_{1}^0} \sim \lambda_{h^0 h^0 h^0}^{SM}$.\end{abstract}

\pacs{12.60.-i,12.60.Fr,14.80.Ec,14.80.Fd,11.30.Er}

\maketitle

\section{Introduction \label{sec:introduction}}
\noindent
The Higgs boson is a fundamental piece of the Standard Model (SM) providing mass to the gauge bosons and fermions upon the spontaneous electroweak symmetry breaking (SSB), and thus preserving the renormalizability of the theory~\cite{Higgs:1964pj,1971NuPhB..35..167T}. In the SM, only one SU(2)${}_L$ doublet Higgs field is included, which upon acquiring a vacuum expectation value breaks the SU(2)${}_L \times$ U(1)${}_Y$ symmetry. Although its existence is a fundamental piece of the theory and the SM Higgs potential is very simple and sufficient to describe a realistic model of mass generation, this may not be the final form of the theory. In the SM, each family of fermions enters independently. To understand the replication of generations and to reduce the number of free parameters, usually more symmetry is introduced in the theory. In this direction interesting work has been done with the addition of discrete symmetries to the SM (see for instance~\cite{Ishimori:2010au,Ishimori:2012zz,Beye:2015wka}  and references therein for a review on the subject).

It is noticeable that many interesting features of masses and mixing of the SM can be understood using a minimal discrete group, namely the permutation group $S(3)$~\cite{Derman:1978iz, Derman:1979nf, Pakvasa:1977in, Pakvasa:1978tx, Mondragon:1998gy, Mondragon:1999jt, Harrison:2003aw, Kubo:2003iw, Kubo:2003pd, Kobayashi:2003fh, Kubo:2004ps, Caravaglios:2005gw, Araki:2005ec, Kubo:2005sr, Koide:2005ep, Grimus:2005mu, Teshima:2005bk, Kimura:2005sx, Koide:2006vs, Mohapatra:2006pu, Kaneko:2007ea,Felix:2006pn,Canales:2013cga,Emmanuel-Costa:2016vej,Vien:2014vka}. 
In the absence of mass, the SM is chiral and invariant with respect to any permutation of the left and right fermionic fields of the same electric charge. For three fermionic families with just one Higgs after SSB, as in the SM, only one quark and one lepton acquire mass. Then, to give mass to all fermions and at the same time preserve the $S(3)$ flavour symmetry of the theory, an extended flavoured Higgs sector is required with three Higgs SU(2) doublets: one in a singlet and the other two in a doublet irreducible representation of $S(3)$~\cite{Kubo:2003iw,Mondragon:2007af,Beltran:2009zz}.

Furthermore, the particle observed at the Large Hadron Collider (LHC) corresponds to the SM physical spectrum. It is not known if there is one or many Higgs bosons, yet an indication of the presence of just one Higgs or an extended Higgs sector, as the one proposed in the $S(3)$-invariant extension of the Standard Model ($S(3)$SM), could be found in a future running at the LHC~\cite {Barger:2009me,Gupta:2009wn}.
Models with more than one Higgs doublet, with or without supersymmetry, have been studied extensively for a review of supersymmetric and two Higgs-doublet models~\cite{Kanemura:2004mg,Djouadi:2005gj,Branco:2011iw}. Different aspects of three and more Higgs doublets models have also been studied, with and without discrete symmetries (see~\cite{Lendvai:1981wn,Adler:1999gv,Ferreira:2008zy,Howl:2009ds}). In particular, in Refs.~\cite{Barroso:2005tq,Barroso:2005da,Barroso:2006pa} it was shown that in two-Higgs doublet models, at tree level, the potential minimum that preserves electric charge and CP symmetries, when it exists, is a stable and global one. Many of these models are not concerned with the unsolved problem of family replication, and thus there is also analysis of different aspects of the Higgs potential of various discrete flavour groups~\cite{Kubo:2004ps, Hagedorn:2006ir, Tofighi:2009zzb, Morisi:2009sc, Morisi:2010rk,EmmanuelCosta:2007zz,Beltran:2009zz}.
A main theoretical goal is to construct a flavoured or extended Higgs potential with SSB in the ground state, which at the same time gives mass to $W^{\pm},\, Z^{0}$ and fermions of the three observed families. The Higgs fields determine the shape of the potential. In this work we consider the symmetry of permutations $S(3)$ where the Higgs sector has three Higgs SU(2) doublets fields~\cite{Beltran:2009zz,EmmanuelCosta:2007zz}. The symmetry $S(3)$ is the smallest non-Abelian discrete group, which offers a possible explanation of why there are three generations of the quarks and leptons~\cite{Mondragon:1998gy}.The Yukawa couplings of the $S(3)$SM are sufficient to reproduce the masses of the quarks and leptons, and can also make predictions in the neutrino sector~\cite{Kubo:2004ps,Dev:2012ns,Dias:2012bh,Canales:2012dr}.

The discovery of a scalar field electrically neutral with a mass of $125.7\pm 0.4$ GeV~\cite{Aad:2013xqa} in the LHC has been done. With the discovery of the Higgs at CERN, July 4, 2012~\cite{Aad:2012tfa,Chatrchyan:2013lba,Chatrchyan:2012ufa}, our understanding of the physics of particles and fields reach a point at which, the SM with one Higgs as a result of the SSB has been confirmed. The next step is setting out the properties of this physical Higgs, mainly its couplings to gauge bosons and fermions, besides its self-couplings~\cite{Miller:2000uz, Baur:2003gp, Dutta:2008bh, Barr:2014sga}.

These properties have to be considered in the analysis of extensions of the SM, whose Higgs sector contains more than one SU(2) doublet Higgs field. So it is crucial to experimentally determine if there is only one or there are more scalar, neutral, or electrically charged Higgs states. As we can see, there are still many unsolved answers, of which, the most important are: Why do we observe the generation's replication? Why do we observe a hierarchy of masses between fermions? Why CP violation? As we know, SSB is the mechanism through which the particles acquire mass, but, which is the reason for the large mass difference between the particles of each generation and why three generations exist as well. 
Moreover, how to explain that neutrinos have a small non-vanishing mass? And where does CP violation come from? These questions remain open. The way we tackled these problems is considering the permutation symmetry $S(3)$, a way to go beyond the SM (BSM)~\cite{Barradas-Guevara:2016ecp,Barradas-Guevara:2014yoa, Bhattacharyya:2010hp, Bhattacharyya:2012ze}. 
Extending the Higgs sector with three SU(2) doublet Higgs fields given an invariant potential under permutation symmetry $S(3)$, one obtains a greater number of physical states of Higgs bosons~\cite{Barradas-Guevara:2014yoa, Das:2014fea}. Moreover, this permutation symmetry allows us to develop exact and analytical solutions for nine physical Higgs bosons in the normal minimum without CP violation as shown in Ref.~\cite{Barradas-Guevara:2014yoa}. In this, we found that the neutral $S(3)$ trilinear Higgs couplings are given by $\lambda_{ijk} = F(\theta_s) \cos\omega_3 + G(\theta_s) \sin\omega_3$, with two mixing angles  $\omega_3=\arctan(2 v_2/v_3)$ and $\theta_s$, among two neutral Higgs bosons $H_{1,2}^0$.
From the numerical analysis, we found a Higgs state $H^0_2$ with a mass of 125 GeV and a trilinear Higgs self-coupling $\lambda_{H^0_2H^0_2H^0_2}$ as the one in the SM~\cite{Barradas-Guevara:2014yoa}. As we know, CP violation is one of the distinctive facts of the electroweak interactions, and CP is a possible symmetry of the electroweak Lagrangians, although it has to be broken. Spontaneous CP violation in the scalar sector has been studied in a lot of works prior to extensions of the SM, see~\cite{Lee:1974jb,Chen:2015gaa} and references therein. In particular, extended scalar sectors show spontaneous CP violation given by a relationship between the vacuum expectation values of the Higgs fields. In this work, we perform a detailed study of the spontaneous CP breaking conditions of $S(3)$SM. This model has been previously used to successfully calculate the Higgs masses spectrum and mixings as well as trilinear Higgs self-couplings~\cite{EmmanuelCosta:2007zz,Beltran:2009zz}, quark and lepton mixing~\cite{Caravaglios:2005gw,Canales:2013cga}, and flavour changing neutral currents (FCNC)~\cite{Kubo:2003iw,Mondragon:2007af}. The model has three $S(3)$ flavoured Higgs fields, $\Phi_{1,2,S}$, which upon acquiring vev's, break the electroweak symmetry. In here, we examine the CP breaking minimization conditions, without explicit breaking of the flavour symmetry, even though it may be spontaneously broken. $S(3)$SM has three different stationary points, which can be classified as Normal, Charge Breaking (CB), and Charge Parity Breaking (CPB) minima, according to the vacuum expectation values of the three Higgs fields~\cite{Beltran:2009zz}. An extended Higgs sector opened up the window for CP violation scenarios coming from it (see section~\ref{sec:mincond}). We found the conditions under which a potential minimum solution reproduces the gauge bosons masses: that is, the CP breaking minimum should be deepest than the normal (N) and charge breaking (CB) stationary points. We described the different CPB scenarios of the model and give expressions for the Higgs mass matrix in section~\ref{sec:higgsmasses}. As we can see in section~\ref{sec:parameterspace}, we ended up with nine Higgs fields. But these physical Higgs states remain to be seen at the LHC; that is, a CP breaking Higgs among $H^{0}_{1,2,4,5}$ could be found at the LHC. A numerical computation of the trilinear Higgs self-couplings $\lambda_{H^0_{i}H^0_{j}H^0_{k}},\, (i,j,k=1,2,4,5)$ allows us to find out $H_4^0$ as the right like-SM Higgs candidate.

\section{The scalar potential in $S(3)$SM \label{sec:potential}} 
\noindent The Lagrangian  ${\cal L}_{H}$ of the  extended Higgs sector $S(3)$SM includes three complex $SU(2)$ doublets fields: 
\be
{\cal L}_{\Phi_i} =\left[   D_\mu \Phi_S\right]^2+\left[D_\mu \Phi_1\right]^2
+\left[D_\mu \Phi_2\right]^2-V\left( \Phi_1,\Phi_2,\Phi_S \right),
\label{eq:one}
\ee
where $D_\mu$ is the usual covariant derivative,
$ D_\mu=\left( \partial_\mu-\frac{i}{2}g_2{\tau_a}  {W_{\mu}^a}-\frac{i}{2}g_1B_\mu \right)$, with $g_1$ and $g_2$ standing for the $U(1)$ and $SU(2)$ coupling constants. 
The most general Higgs potential $V\left( \Phi_1,\Phi_2,\Phi_S \right)$
invariant under $SU(3)_C \times SU(2)_L \times U(1)_Y\times S(3)$ can be written 
as~\cite{EmmanuelCosta:2007zz,Beltran:2009zz}:
\be \label{eq:potential}
\ba{rcl}
V\left( \Phi_1,\Phi_2,\Phi_S \right)&=&\mu_1^2\left(\Phi^\dagger_1 \Phi_1+ \Phi^\dagger_2 \Phi_2\right)+ \mu_0^2\left(\Phi^\dagger_S \Phi_S\right)+
a\left(\Phi^\dagger_S \Phi_S\right)^2
+ b\left( \Phi^\dagger_S \Phi_S \right)\left( \Phi^\dagger_1 \Phi_1+\Phi^\dagger_2 \Phi_2 \right) \\
&+&
 c\left( \Phi^\dagger_1 \Phi_1+\Phi^\dagger_2 \Phi_2 \right)^2
+
d\left(\Phi^\dagger_1 \Phi_2-\Phi^\dagger_2 \Phi_1\right)^2+ e{\it  f}_{ijk}\left(\left(\Phi^\dagger_S \Phi_i\right)\left(\Phi^\dagger_j \Phi_k\right)+\textrm{H.C.}\right)
\\
 &+& 
 f\left\{\left(\Phi^\dagger_S \Phi_1\right)\left(\Phi^\dagger_1 \Phi_S\right)+\left(\Phi^\dagger_S \Phi_2\right)
\left(\Phi^\dagger_2 \Phi_S\right) \right\}
+g\left\{ \left(\Phi^\dagger_1 \Phi_1-\Phi^\dagger_2 \Phi_2\right) ^2+
\left(\Phi^\dagger_1 \Phi_2+\Phi^\dagger_2 \Phi_1\right)^2\right\}
\\
& +&h\bigg\{\left(\Phi^\dagger_S \Phi_1\right)
\left(\Phi^\dagger_S \Phi_1\right)+\left(\Phi^\dagger_S \Phi_2\right)\left(\Phi^\dagger_S
\Phi_2\right)+\left(\Phi^\dagger_1 \Phi_S\right)\left(\Phi^\dagger_1 \Phi_S\right) + \left(\Phi^\dagger_2 \Phi_S\right)\left(\Phi^\dagger_2 \Phi_S\right)\bigg\},
\ea
\ee
where $f_{112}=f_{121}=f_{211}=-f_{222}=1$, and $\mu_0^2$, $\mu_1^2$ are mass parameters; $a$, $b$, $\cdots$, $h$ are real and dimensionless parameters.  We can  write down the $SU(2)$ Higgs  doublets to include the discrete flavour symmetry $S(3)$  as 
\be
\ba{rcl} \label{eq:doubletshiggs}
\Phi_1&=&\left( \ba{c}
\phi_1+i\phi_4\cr
\phi_7+i\phi_{10}
\ea
\right),\,
\Phi_2=\left( \ba{c}
\phi_2+i\phi_5\cr
\phi_8+i\phi_{11}\ea
\right), \,
\Phi_S=\left( \ba{c}
\phi_3+i\phi_6\cr
\phi_9+i\phi_{12}\ea
\right) .
\ea
\ee
The numbering of the real scalar $\phi_i$ fields is chosen for convenience when writing the mass matrices for the scalar particles, and the subscript $S\equiv 3$ is the flavour index for the Higgs  singlet field under $S(3)$. $\Phi_i$ ($i=1,2$) are the components of the $S(3)$ doublet field. 
In the analysis, it is better to introduce nine real quadratic forms $x_i$ invariant under $SU(2)\times U(1)$ given as
\be\label{eq:xveccomp}
\ba{ccc}
x_1=\Phi^\dagger_1 \Phi_1,&
x_4= {\cal R}\left(\Phi^\dagger_1 \Phi_2\right),&
x_7= {\cal I}\left(\Phi^\dagger_1 \Phi_2\right),\cr
x_2=\Phi^\dagger_2 \Phi_2,&
x_5= {\cal R}\left(\Phi^\dagger_1 \Phi_S\right),&
x_8= {\cal I}\left(\Phi^\dagger_1 \Phi_S\right),\cr
x_3=\Phi^\dagger_S \Phi_S ,&
x_6= {\cal R}\left(\Phi^\dagger_2 \Phi_S\right),&
x_9={\cal I}\left(\Phi^\dagger_2 \Phi_S\right) .
\ea
\ee
Now, it is a simple matter to write down the $S(3)$SM potential (\ref{eq:potential}),
\be \label{eq:potential1}
\ba{lll}
V(x_1,\cdots,x_9)&=&\mu^2_1\left(x_1+x_2 \right)+\mu^2_0 x_3+ax^2_3+b\left(x_1+x_2 \right)x_3+c\left(x_1+x_2 \right)^2\\[.3cm]
&&-4dx_7^2+2e \left[ \left(x_1 - x_2\right)x_6+2x_4 x_5\right]+f\left(x_5^2+x_6^2+x_8^2+x_9^2 \right) \\[.3cm]
&&+g\left[\left(x_1-x_2 \right)^2+4x_4^2 \right]+2h\left(x_5^2+x_6^2-x_8^2-x_9^2 \right) ;\\
\ea
\ee
and we can  rewrite the potential $V(x_1,\cdots,x_9)$ and express it in a simple matrix form as
\be \label{eq:potential2}
V({\bf X} )={\bf A}^T{\bf X}+\frac{1}{2}{\bf X}^T{\bf B}{\bf X}.
\ee
The vector  ${\bf X}$ given by 

\be
{\bf X}^T=\left( x_1, x_2 ,x_3, \dots,x_9 \right),\label{appppa} 
\ee
${\bf A}$ is a mass parameter vector
\begin{eqnarray}\label{eq:amatrix}
 {\bf A}^T=\left(\mu^2_1,\mu^2_1,\mu^2_0,0,0,0,0,0,0 \right)  \label{appppb}
\end{eqnarray}
and ${\bf B}$  is a $9\times9$ real parameter symmetric matrix
\be \label{eq:bmatrix}
{\bf B}= \left( \ba{ccccccccc}
2(c+g) & 2(c-g) &b&0&0&2e&0&0&0\cr
2(c-g) & 2(c+g) & b&0&0&-2e&0&0&0\cr
b&b&2a&0&0&0&0&0&0\cr
0&0&0&8g&4e&0&0&0&0\cr
0&0&0&4e&2(f+2h)&0&0&0&0\cr
2e&-2e&0&0&0&2(f+2h)&0&0&0\cr
0&0&0&0&0&0&-8d&0&0\cr
0&0&0&0&0&0&0&2(f-2h)&0\cr
0&0&0&0&0&0&0&0&2(f-2h)
\ea
\right).
\ee
The matrix ${\bf B}$ must be positive definite~\cite{VALIAHO198619,Unwin:2011rn}, as this is fundamental to study the critical points in the Higgs potential. Then it is quite straightforward to find the following necessary conditions for the global stability in the asymptotic limit:
$$
a, f, g > 0 ,\quad c \geq \frac{b^2}{4a} , \quad d, e  < 0 , \quad 
\frac{e^2-fg}{2g} < h < \frac{f}{2} .
$$

 In the CP conserving case, the vacuum expectation values of the Higgs doublets are taken as real values. This case was carried out in Ref.~\cite{Barradas-Guevara:2014yoa}, which was considered as the normal minimum with
$$
\phi_7={\it v_1},\ \phi_8={\it v_2},\ \phi_9={\it v_3},\
\phi_i=0, \,\,\,\, i\neq 7,8,9\, ,
$$ 
 where we have adopted for convenience vev's $v_i$ ($i=1,2,3$), 
with $v_i \in \Re$. The CP breaking minimum (CPB) we have%
\be
\langle \Phi_i \rangle = \displaystyle\frac{1}{\sqrt{2}} \left( \begin{array}{c} 0 \\ v_i + i \gamma_i \end{array}\right) \, \qquad
i = 1, 2, 3 \, ,
\ee
where $\gamma_i \in \Re$. Then, CPB is at
\be\label{eq:minimcpviolation}
\begin{array}{c}
\phi_7=v_1,\ \phi_8=v_2,\ \phi_9=v_3, \
\phi_{10}= \gamma_1,\ \phi_{11}=\gamma_2,\ \phi_{12}=\gamma_3,\qquad \hbox{and other cases} \qquad \phi_i=0 \, ,
\end{array}
\ee 
which should satisfy the constraint
\be
v= \left(v_1^2 + v_2^2 + v_3^2+ \gamma_1^2 + \gamma_2^2 + \gamma_3^2\right)^{1/2}. \label{eq:constraint}
\ee

To complete the story, the constants $\gamma_i$ can take the values following:
\begin{itemize}
\item
$ \gamma_1 \neq 0$, and $\gamma_2 = \gamma_3 = 0$;
\item
$ \gamma_2 \neq 0$, and $\gamma_1 = \gamma_3 = 0$;
\item
$ \gamma_3 \neq 0$, and $\gamma_1 = \gamma_2 = 0$;
\item
$ \gamma_1 \neq 0$, $\gamma_2 \neq 0$, and $\gamma_3 = 0$;
\item
$ \gamma_1 \neq 0$, $\gamma_3 \neq 0$, and $\gamma_2 = 0$;
\item
$ \gamma_2 \neq 0$, $\gamma_3 \neq 0$, and $\gamma_1 = 0$; and
\item
$ \gamma_1 \neq 0$, $\gamma_2 \neq 0$, and $\gamma_3 \neq 0$.
\end{itemize}
We assume the Higgs vev's are free parameters subject to the constraint~(\ref{eq:constraint}). 

The potential parameters in eq.~(\ref{eq:potential}), specifically the mass parameters $\mu_0^2$ and $\mu_1^2$, may be written in terms of the vev's. The fermions in the $S(3)$SM acquire mass through the Yukawa interactions~\cite{Kubo:2003iw}, but once the Higgs fields break the gauge symmetry, all fermions acquire mass. The Yukawa couplings may all be complex, particularly those with real values as their corresponding Yukawa Lagrangian is given in Ref.~\cite{Kubo:2003iw}. From this, we can express the fermionic mass matrix ${\bf M}_f$ including spontaneous CP violation ($\gamma_3 \neq 0$ and $\gamma_1=\gamma_2=0$) as
\begin{equation}
{\bf M}_f=\left( 
\begin{array}{ccc}
m_{1}^{CP}+m_{6} & m_{2}^{{}} & m_{5}^{{}} \\ 
m_{2}^{{}} & m_{1}^{CP}-m_{6} & m_{8} \\ 
m_{4}^{{}} & m_{7} & m_{3}^{CP}
\end{array}
\right) ,
\end{equation}
where 
\begin{eqnarray}
m_{1}^{CP} &=&m_{1} -Y_{1}^{f}\left( i\gamma_3 \right), \\
m_{3}^{CP} &=&m_{3} -Y_{3}^{f}\left(i\gamma_3 \right). 
\end{eqnarray}
$m_{i} \, (i=1,2,\cdots,8$) are the expressions in the case of CP conserving~\cite{Kubo:2003iw}.  Then, the fermionic mass matrices are complex caused by contribution $\gamma_3$ arising from the Higgs sector. Thus, the SSB mechanism provides a source for CP violation in the fermionic sector and contributes to the same in the quark and lepton mixing matrices. 

\section{Minimum conditions \label{sec:mincond}}
\noindent In this section, we present the minimum conditions and the parameter space analysis for each considered scenario. The minimization conditions give us six equations  determined by demanding of $\partial V/\partial \phi_i\mid_{{min}}=0$.
We denote $M_i{(\gamma_1,\gamma_2,\gamma_3)} \equiv \partial V/\partial \phi_i\mid_{min}$.

\subsection{Scenario 1: $\gamma_1 \neq 0$ and $\gamma_2=\gamma_3=0$ \label{subsec:scen1}}
For this scenario, we have
\begin{subequations}
\begin{alignat}{6}
M_7({\gamma_1})    =&\displaystyle\frac{v_1}{\sqrt{2}} \left[v_3^2 k_2+2 \left(\gamma _1^2 k_1+\mu _1^2\right)+2 \left(v_1^2+v_2^2\right) k_1+6 e v_2 v_3\right]  \label{eq:condmincase1a},\\
M_8({\gamma_1}) =& \displaystyle\frac{v_2}{\sqrt{2}}\left[v_3^2 k_2+2 \left(\gamma _1^2 k_3+\mu _1^2\right)+2\left(v_1^2 +v_2^2\right)  k_1  +\displaystyle\frac{ e v_3}{v_2}\left(3(v_1^2-v_2^2)+\gamma_1^2\right)\right],\label{eq:condmincase1b}\\
M_9({\gamma_1}) =& \displaystyle\frac{v_3}{\sqrt{2}}\left[2 a v_3^2+\gamma _1^2 k_2'+2 \mu _0^2
+\left(v_1^2+v_2^2\right) k_2+\displaystyle\frac{ e v_2}{v_3}\left(3v_1^2-v_2^2+\gamma_1^2\right)\right], \label{eq:condmincase1c}\\
M_{10}({\gamma_1}) =&  \displaystyle\frac{\gamma _1}{\sqrt{2}} \left[v_3^2 k_2'+2 v_2^2 k_3+2 \left(\gamma _1^2 k_1+\mu _1^2\right)+2 v_1^2 k_1+2 e v_2 v_3\right],\label{eq:condmincase1d}\\
M_{11}({\gamma_1}) =&\sqrt{2} \gamma _1 v_1 \left(2 v_2 k_4+e v_3\right),\label{eq:condmincase1e}\\
M_{12}({\gamma_1}) =& \sqrt{2} \gamma _1 v_1 \left(e v_2+2h v_3\right),\label{eq:condmincase1f}
\end{alignat}
\end{subequations}
where we adopt the abbreviations 
\be
\begin{array}{ll}
k_1=c+g, &
k_2=b+f+2h, \\ \\
k_2'=b+f-2h, &
k_3=c-2d-g, \\ \\
k_4=d+g.
\end{array}
\ee
Then, following our earlier analysis, we would have $\mu_1^2$ and $\mu_0^2$ as

\begin{eqnarray}
\mu_1^2 &=& v_2^2 k_6-\displaystyle\frac{1}{2} v_3^2 k_5,  \label{eq:termmassg1a} \\ 
\mu_0^2 &=& -a v_3^2-2 v_2^2 k_5+\displaystyle\frac{4 v_2^4 k_4}{v_3^2},  \label{eq:termmassg1b}
\end{eqnarray}
They are independent of CPB vev $\gamma_1$; here we have used the abbreviations
\be
\begin{array}{l}
k_5 = b+f \\ 
k_6 = -4c + 5d + g.
\end{array}
\ee
We can obtain the free parameters $e$ and $h$ from eq.~(\ref{eq:condmincase1e}) and eq.~(\ref{eq:condmincase1f}) respectively:
\begin{equation}
\frac{e}{h} = -\displaystyle\frac{2 v_3 }{v_2}. \label{fraceh}
\end{equation}
Next, using eq.~(\ref{eq:condmincase1a}) and eq.~(\ref{eq:condmincase1b}) we obtain the possible solution
\be
 v_1=
   \pm\sqrt{3 v_2^2-\gamma
   _1^2}.\label{eq:v1g1}
\ee
Thereby, the mass parameters $\mu_1^2$~(\ref{eq:termmassg1a}) and $\mu_0^2$~(\ref{eq:termmassg1b}), a dimensionless parameter $e/h$, eq.~(\ref{fraceh}), are functions of the vacuum expectation values $v_2$, $v_3$.  
This scenario is interesting because, from twelve degrees of freedom after SSB is done, eight physical Higgs and four Goldstone bosons were obtained. Later, we will discuss this further.

\subsection{Scenario 2: $\gamma_2 \neq 0$ and $\gamma_1=\gamma_3=0$ \label{subsec:scen2}}
As we derived in the previous scenario, to determine the model restrictions, again minimizing the potential we obtain the constraints as follows
\begin{subequations}
\begin{alignat}{6}
M_7({\gamma_2}) =& \displaystyle\frac{v_1}{\sqrt{2}}\left[ v_3^2 k_2+2
   \left(\gamma _2^2 k_3+\mu
   _1^2\right)  +2( v_1^2 + v_2^2) k_1+6
   e v_2 v_3\right]\label{eq:condmin2a},\\
M_8({\gamma_2}) =& \displaystyle\frac{v_2}{\sqrt{2}}\left[ v_3^2 k_2+2
   \left(\gamma _2^2 k_1+\mu
   _1^2\right)  +2\left(v_1^2 +v_2^2\right) 
   k_1+\displaystyle\frac{e v_3}{v2}\left(3 \left(v_1^2+v_2^2\right) -\gamma _2^2\right) \right]\label{eq:condmin2b},\\
M_9({\gamma_2}) =& \displaystyle\frac{v_3}{\sqrt{2}}\left[2 a v_3^2+\gamma _2^2 k_2'+2 \mu _0^2 +
   k_2(v_1^2 +v_2^2) +\displaystyle\frac{e v_2}{v_3}( 3 v_1^2-v_2^2-\gamma _2^2) \right]\label{eq:condmin2c},\\
M_{10}({\gamma_2}) =& \sqrt{2} v_1 \gamma _2 \left(2 v_2 k_4+e
   v_3\right) \label{eq:condmin2c},\\
M_{11}({\gamma_2}) =& \displaystyle\frac{\gamma _2}{\sqrt{2}} \left[v_3^2 k_2'+2
   v_1^2 k_3 +2  k_1(v_2^2+\gamma _2^2)
+2\mu _1^2-2
   e v_2 v_3\right] \label{eq:condmin2d},\\
M_{12}({\gamma_2}) =& \displaystyle\frac{\gamma _2}{\sqrt{2}} \left[e (v_1^2- v_2^2-
   \gamma _2^2)+4 h v_2 v_3\right] \label{eq:condmin2f}.
\end{alignat}
\end{subequations}
From eqs.~(\ref{eq:condmin2c}) and~(\ref{eq:condmin2f}), we obtain the  parameters $e$ and $h$:
\begin{equation}
\frac{e}{h} = -\displaystyle\frac{2 v_3 }{v_2}. 
\end{equation}
Using eqs.~(\ref{eq:condmin2a}) and~(\ref{eq:condmin2b})
\be
v_1 = \pm \sqrt{3
   v_2^2+\gamma _2^2}.
\ee
Therefore,
\be
\begin{array}{l}
\mu_1^2 = \displaystyle\frac{1}{2}\left[-v_3^2 k_5+2 v_2^2 k_6+4 \gamma
   _2^2 k_7\right],
\\
\mu_0^2 = -a v_3^2-2 v_2^2k_5 +\displaystyle\frac{4
   v_2^4 k_4}{v_3^2}.
   \end{array}
   \ee
Unlike scenario 1,  $\mu_1^2$ has a dependence on the CPB parameter $\gamma_2$, here $k_7 = d-c$, but we should also have an acceptable Higgs masses set. As in the previous scenario, we obtain eight physical Higgs fields and four Goldstone bosons.

\subsection{Scenario 3: $\gamma_3 \neq 0$ and $\gamma_1=\gamma_2=0$ \label{subsec:scen3}}
In this scenario, the equations that result from the CPB minimum conditions are
\begin{subequations}
\begin{alignat}{2}
M_{7}(\gamma_3) &= \displaystyle\frac{v_1}{\sqrt{2}} \left[\left(\gamma _3^2+v_3^2\right)k_5
   -2 h(\gamma _3^2-v_3^2)   +2(
   v_1^2 + v_2^2) k_1+6
   e v_2 v_3+2 \mu
   _1^2\right]\label{eq:condmin3a},\\
M_{8}(\gamma_3) &= \displaystyle\frac{v_2}{\sqrt{2}} \left[\left(\gamma _3^2+v_3^2\right)
   k_5-2h\left(\gamma _3^2-v_3^2\right) +2\left(v_1^2
  +v_2^2\right) k_1+2
   \mu _1^2+\displaystyle\frac{3 e
   v_3}{v_2}
   \left(v_1^2-v_2^2\right)\right] \label{eq:condmin3b},
\\
M_{9}(\gamma_3) &= \displaystyle\frac{v_3}{\sqrt{2}}\left[2  \left(a
   \left(\gamma
   _3^2+v_3^2\right)+\mu
   _0^2\right) +\left(v_1^2+v_2^2\right) k_2+ \displaystyle\frac{e v_2}{v_3}\left(3 v_1^2-v_2^2\right)\right] \label{eq:condmin3c},
\\
M_{10}(\gamma_3) &= \sqrt{2} \gamma _3 v_1 \left(e
   v_2+2 h v_3\right) \label{eq:condmin3d},
\\
M_{11}(\gamma_3) &= \displaystyle\frac{\gamma _3}{\sqrt{2}} \left[v_2
   \left(4 h v_3-e
   v_2\right)+e
   v_1^2\right] \label{eq:condmin3e},
\\
M_{12}(\gamma_3) &= \displaystyle\frac{\gamma _3}{\sqrt{2}} \left[2
   \left(a \left(\gamma
   _3^2+v_3^2\right)+\mu
   _0^2\right)+\left(v_1^2+v_2
   ^2\right) (b+f-2
   h)\right].\label{eq:condmin3f}
\end{alignat}
\end{subequations}
Using eq.~(\ref{eq:condmin3a}) we have 
\begin{eqnarray}
\mu_1^2  &=& \displaystyle\frac{1}{\sqrt{2}}\left(-\gamma _3^2
   k_2'-v_3^2 (b+f-10
   h)-8 v_2^2
   k_1\right),\label{eq:valmu1c3}
   \\
\mu_0^2  &=& -a  \left(\gamma
   _3^2+v_3^2\right)-2 
   v_2^2 (b+f-4 h)+\displaystyle\frac{2 e
   v_2^3}{v_3}. \label{eq:valmu0c3}
\end{eqnarray}

From eqs.~(\ref{eq:condmin3d}) and~(\ref{eq:condmin3e})
\begin{equation}
\frac{e}{h} = -\displaystyle\frac{2 v_3 }{v_2}. 
\end{equation}
and using eqs.~(\ref{eq:condmin3a}) and~(\ref{eq:condmin3b}), we obtain 
\be v_1=\sqrt{3}v_2,
\label{eq:v1eq3v2}
\ee
as in the normal minimum, 
unlike scenario 1,  $\mu_0^2$ and $\mu_1^2$ has a dependence on the CPB parameter $\gamma_3$.
We showed the results for different scenarios where CP-violation was realized. In each scenario, we computed the Higgs mass matrix and Higgs mass eigenvalues as follows.
\begin{center}
\begin{table}
\begin{tabular}   {   |c||c||c||c|   } \hline
\multicolumn{4}{|c|}{\bf Comparison of the potential variables in the three scenarios} \\  \hline\hline
\multicolumn{1}{|c||}{\bf Parameter} & \multicolumn{1}{|c||}{\bf Scenario 1} & \multicolumn{1}{c||}{\bf Scenario 2} & \multicolumn{1}{c|}{\bf Scenario 3} \\  \hline
& & & \\
$\displaystyle\frac{e}{h}$ &  $-\displaystyle\frac{2 v_3}{v_2}$ & $-\displaystyle\frac{2 v_3 }{v_2}$ & $-\displaystyle\frac{2  v_3}{v_2}$ \\  
& & & \\ \hline
\hline
\hline
\multicolumn{4}{|c|}{\bf The vacuum expectation value (vev)} \\  \hline\hline
& & & \\
$v_1$ & $\sqrt{3 v_2^2-\gamma_1^2}$  & $\sqrt{3v_2^2+\gamma _2^2}$  & $\sqrt{3}v_2$ \\ [.3cm]
\hline
& & & \\
$v$ & $\sqrt{4 v_2^2+v_3^2}$  & $\sqrt{4v_2^2+v_3^2+2\gamma _2^2}$  & $\sqrt{4v_2^2+v_3^2+\gamma_3^2}$ \\ [.3cm]
\hline
\hline
\multicolumn{4}{|c|}{\bf Mass terms} \\  \hline\hline
& & & \\
$\mu_1^2$ & $\begin{array}{c}v_2^2 k_6\\ -\displaystyle\frac{1}{2} v_3^2 k_5\end{array}$  & $\begin{array}{c}-\displaystyle\frac{1}{2}v_3^2 k_5 \\ + v_2^2 k_6\\+2 \gamma
   _2^2 k_7\end{array}$  & $\begin{array}{c}\displaystyle\frac{1}{\sqrt{2}}\left(-\gamma _3^2
   (b+f-2h)\right. \\ -v_3^2 (b+f-10
   h) \\ \left. -8 v_2^2 (c+g)\right)\end{array}$ \\ [.9cm]
\hline
& & & \\
$\mu_0^2$ & $\begin{array}{c} -a v_3^2-2 v_2^2 k_5\\ +\displaystyle\frac{4 v_2^4 k_4}{v_3^2}\end{array}$  & $\begin{array}{c}-a v_3^2-2 v_2^2k_5 \\ + \displaystyle\frac{4
   v_2^4 k_4}{v_3^2}\end{array}$  & $\begin{array}{c}-a  \left(\gamma
   _3^2+v_3^2\right) \\ -2 
   v_2^2 (b+f-4 h)\\ +\displaystyle\frac{2 e
   v_2^3}{v_3}\end{array}$ \\ [1cm]
\hline
\hline
\multicolumn{4}{|c|}{ } \\ 
\multicolumn{4}{|c|}{ $k_4= d+g$, $k_5 = b+f$, $k_6 = -4c+5d+g$, $k_7 = d-c$. } \\ 
\multicolumn{4}{|c|}{ } \\ \hline
\end{tabular}
\caption{Relationships of the three CPB scenarios.}
\end{table}
\end{center}
%

\section{Higgs masses \label{sec:higgsmasses}}
The Higgs mass matrix is obtained from the computation of the second derivatives of the Higgs potential, 
eq.~(\ref{eq:potential}). There are twelve real Higgs fields $\phi_i$, and the corresponding Higgs mass matrix is a 12 $\times$ 12 real matrix, then 
\be
(\mathcal{M}^2_H)_{ij} = \displaystyle\left.\displaystyle\frac{1}{2}\displaystyle\frac{\partial^2 V}{\partial\phi_i\partial\phi_j}\right|_{\hbox{CPBmin}} ,
\ee
with $i,j={1,2,....,12}$. We have
\be\label{eq:matrizmasa}
\mathcal{ M}^2_H = \hbox{diag}\left({\bf M}_{C,\gamma}^2, {\bf M}_{N,\gamma}^2   \right) \, ,
\ee
with ${\bf M}_{C,\gamma}^2$ corresponding to the mass matrix of electrically charged Higgs bosons and ${\bf M}_{N,\gamma}^2$ to the neutral Higgs mass matrix, which are
the $6\times 6$ symmetric and Hermitian sub-matrices. 

For each of the corresponding scenarios, we have a matrix for charged and neutral Higgs bosons, that we specify with the gamma index, $\gamma = \gamma_1,\gamma_2,\gamma_3$, as the corresponding scenario, where

\be
{\bf M}^2_{C, \gamma} = \left(
\begin{array}{cc}
{\bf {M}^2_C}_{11}(\gamma) & {\bf {M}^2_C}_{12}(\gamma)  \\ [.4cm]
{\bf {M}^2_C}_{21}(\gamma) & {\bf {M}^2_C}_{22}(\gamma) 
\end{array} \right), \label{eq:matrizcargada}
\ee
which should satisfy the constraint 
\be
\begin{array}{c}
 {\bf {M}^2_C}_{22}(\gamma) = {\bf {M}^2_C}_{11}(\gamma), \\[.4cm]
{\bf {M}^2_C}_{21}(\gamma) = - {\bf {M}^2_C}_{12}(\gamma) .
\end{array} \label{eq:masascargadas}
\ee

The neutral Higgs mass matrix is given by
\be
{\bf M}^2_{N,\gamma} = \left(
\begin{array}{cc}
{\bf {M}^2_N}_{11}(\gamma) & {\bf {M}^2_N}_{12}(\gamma) \\[.4cm] 
{\bf {M}^2_N}_{21}(\gamma) & {\bf {M}^2_N}_{22}(\gamma) 
\end{array} \right), \label{eq:matrizneutra}
\ee
with
\be
\begin{array}{c}
 {\bf {M}^2_N}_{22}(\gamma) \neq {\bf {M}^2_N}_{11}(\gamma), \\[.4cm]
{\bf {M}^2_N}_{21}(\gamma) = {\bf {M}^2_N}_{12}^{\rm T}(\gamma). 
\end{array} \label{eq:masasneutras}
\ee
Here, ${\bf {M}^2_N}_{12}^{\rm T}(\gamma)$ is the transposed matrix of ${\bf {M}^2_N}_{12}(\gamma)$. 
For the three scenarios, the restrictions ~(\ref{eq:masascargadas}) and~(\ref{eq:masasneutras}) were met. The Higgs masses are obtained by diagonalizing the matrices~(\ref{eq:matrizcargada}) and~(\ref{eq:matrizneutra}), for each of the scenarios. How can we know which scenario has got a physically possible situation? We calculated the eigenvalues for the matrices of each scenario. In appendices~\ref{apendiceA} and~\ref{apendiceB}, we show the calculations: we got four null Higgs mass eigenvalues for scenarios 1 and 2 and just three null Higgs mass eigenvalues for scenario 3. Then, we compared that to the Higgs masses and trilinear Higgs self-couplings numerical analysis.

For this model with CP violation arising from the Higgs $S(3)$ doublet sector, among the nine physical Higgs fields, we have four charged bosons which are mass degenerate two by two and four non-degenerated bosons in the neutral sector (see appendices~\ref{apendiceA} and~\ref{apendiceB}). Nevertheless, when CP breaking arises from the $S(3)$ Higgs singlet, we found a physical scenario with three Goldstone bosons, which can give mass to vector bosons $W^{\pm}$ and $Z^0$, with a massless photon and nine physical Higgs fields. At least one neutral Higgs should have a mass of 125.7 $\pm$ 0.4 GeV while the remaining eight additional Higgs states are candidates for new particles. This scenario provides a strong motivation to extend the analysis to CPB phenomenology arising from spontaneous electroweak symmetry breaking. We denote the masses of these Higgs charged bosons as $M_{C_i}$ and $M_{H_j^0}$ for the neutral masses, where $i=1,2$ and $j=1,\cdots,5$. In the following section, we analyze the Higgs masses only for scenario 3.

\section{Parameter space \label{sec:parameterspace}}
In this section, we explore parameter space regions where the model is consistent.  The allowed parameter space is that the Higgs masses are positive~\cite{VALIAHO198619,Unwin:2011rn}. From  eq.~(\ref{eq:v1eq3v2}),  $v_2$ and $v_1$ are expressed in terms of $v_3$ 
and from eq.~(\ref{eq:constraint}),  we found
$$
v^2 = (1+16\frac{h^2}{e^2})v_3^2 + \gamma_3^2.
$$
Hence,  we have defined $\tan \omega$ as
\begin{equation}\label{eq:ratiovevs}
\tan \omega \equiv \frac{\gamma_3}{\alpha v_3}, \quad  \alpha = \sqrt{1+16 h^2/e^2},
\end{equation}
where $\omega \in \left(-\pi, \pi\right)$. 

The mass squared matrix of the charged Higgs is given by the $3 \times 3$ matrix 
$$
M_{\rm C}^2(\gamma_3) = 
\left(
\begin{array}{ccc}
 -\frac{\left((f-6 h) e^2+16 g h^2\right) v_3^2+e^2 (f-2 h)
   \gamma _3^2}{e^2} & -\frac{4 \sqrt{3} h \left(e^2-4 g
   h\right) v_3^2}{e^2} & -\frac{2 \sqrt{3} (f-2 h) h v_3
   \left(v_3-i \gamma _3\right)}{e} \\
 -\frac{4 \sqrt{3} h \left(e^2-4 g h\right) v_3^2}{e^2} &
   -\frac{\left((f-14 h) e^2+48 g h^2\right) v_3^2+e^2 (f-2
   h) \gamma _3^2}{e^2} & -\frac{2 (f-2 h) h v_3 \left(v_3-i
   \gamma _3\right)}{e} \\
 -\frac{2 \sqrt{3} (f-2 h) h v_3 \left(v_3+i \gamma
   _3\right)}{e} & -\frac{2 (f-2 h) h v_3 \left(v_3+i \gamma
   _3\right)}{e} & \frac{16 h^2 (2 h-f) v_3^2}{e^2} \\
\end{array}
\right).
$$
$M_{H_i^\pm}^2$ ($i=0,1,2$) are the charged Higgs mass eigenstates of $M_{\rm C}^2(\gamma_3)$ expressed as
\begin{equation}
\begin{array}{l}
M_{H_0^\pm}^2 = 0, \\
M_{H_1^\pm}^2 = -\displaystyle\frac{(f-2 h) \left(\gamma _3^2 e^2+v_3^2
   \left(e^2+16 h^2\right)\right)}{e^2}, \\
M_{H_2^\pm}^2=  -\displaystyle\frac{v_3^2
   \left(e^2 (f-18 h)+64 g h^2\right)+\gamma _3^2 e^2 (f-2
   h)}{e^2}.
\end{array}
\end{equation}

The minimum we are working with is a global one and hence stable. Then, for $M_{H_i^\pm}^2>0$ ($i=0,1,2$) is necessary and sufficient that $2h \geq f\, , \,  {\rm and} \, f,\ g,\ h > 0$.  Hence, neutral Higgs mass matrix 
\begin{equation} \label{eq:matrixneutra}
M_{{\rm N},\gamma_3}^2= \left(
\begin{array}{cc} 
M_{{\rm N}_{11}}^2 & M_{{\rm N}_{12}}^2 \\
M_{{\rm N}_{21}}^2 & M_{{\rm N}_{22}}^2 
\end{array} \right),
\end{equation} 
where $M_{{\rm N}}^2(\gamma_3) =\left( M_{{\rm N},\gamma_3}^2\right)^{\rm T} \in \mathbb{Re}^{6\times 6}$ is copositive, then $\left( M_{{\rm N},\gamma_3}^2\right)_{ii} \geq 0$ for all $i$.  Then equations (C4), (C5) and (C6) are transformed into
$$
M_{{\rm N}_{11}}^2(\gamma_3) =
\left(
\begin{array}{ccc}
 \frac{48 (c+g) h^2 v_3^2}{e^2} & \frac{4 \sqrt{3} h \left(4
   (c+g) h-3 e^2\right) v_3^2}{e^2} & -\frac{4 \sqrt{3}
   (b+f-4 h) h v_3^2}{e} \\
 \frac{4 \sqrt{3} h \left(4 (c+g) h-3 e^2\right) v_3^2}{e^2}
   & \frac{8 h \left(3 e^2+2 (c+g) h\right) v_3^2}{e^2} &
   -\frac{4 (b+f-4 h) h v_3^2}{e} \\
 -\frac{4 \sqrt{3} (b+f-4 h) h v_3^2}{e} & -\frac{4 (b+f-4 h)
   h v_3^2}{e} & \frac{4 \left(16 h^3+a e^2\right)
   v_3^2}{e^2} \\
\end{array}
\right),
$$
$$
M_{{\rm N}_{12}}^2(\gamma_3) =
\left(
\begin{array}{ccc}
 0 & -4 \sqrt{3} h v_3 \gamma _3 & -\frac{4 \sqrt{3} (b+f-2
   h) h v_3 \gamma _3}{e} \\
 -4 \sqrt{3} h v_3 \gamma _3 & 8 h v_3 \gamma _3 & -\frac{4
   (b+f-2 h) h v_3 \gamma _3}{e} \\
 -\frac{8 \sqrt{3} h^2 v_3 \gamma _3}{e} & -\frac{8 h^2 v_3
   \gamma _3}{e} & 4 a v_3 \gamma _3 \\
\end{array}
\right),
$$
$$
M_{{\rm N}_{22}}^2(\gamma_3) =
\left(
\begin{array}{ccc}
 \frac{4 h \left(\left(e^2-4 (d+g) h\right) v_3^2+e^2 \gamma
   _3^2\right)}{e^2} & \frac{4 \sqrt{3} h \left(4 (d+g)
   h-e^2\right) v_3^2}{e^2} & 0 \\
 \frac{4 \sqrt{3} h \left(4 (d+g) h-e^2\right) v_3^2}{e^2} &
   \frac{4 h \left(3 \left(e^2-4 (d+g) h\right) v_3^2+e^2
   \gamma _3^2\right)}{e^2} & 0 \\
 0 & 0 & 4 a \gamma _3^2 \\
\end{array}
\right) .
$$
The neutral Higgs mass eigenstates of $M_{{\rm N},\gamma_3}^2$ matrix are $M^2_{H^0_0}$, $M^2_{H^0_1}$, $M^2_{H^0_2}$, $M^2_{H^0_3}$, $M^2_{H^0_4}$, and $M^2_{H^0_5}$  of which the first is zero, as noted in the aforementioned section. We noticed that, in general, there exist multiple minima in the 3HDM potential. However, with our choice of input parameters including Higgs squared masses and these being positive, we assume that the potential (\ref{eq:potential}) is bounded from below, which happens iff
\begin{equation}
a, b, c, f, g, h \geq 0 \quad {\rm and} \quad e, d \leq 0.
\end{equation}
The lower mass values of the neutral Higgs, nonzero specifically, correspond to $M_{H_1^0}$, and $M_{H_4^0}$, while higher Higgs mass values are allowed for $M_{H_{2,3,5}^0}$, reaching values greater than 1 TeV. Therefore, this model $S(3)$SM has eight free parameters: seven Higgs masses ($M_{H_{1,2,3,4,5}^0}$, and $M_{H_{1,2}^\pm}$), and the ratio of $\gamma_3$ between $\alpha v_3$, $\tan \omega$,  eq.~(\ref{eq:ratiovevs}). In our numerical analysis, the values of the quartic parameters are set, such that they secure the masses of the Higgses, and only consider $\tan \omega$. It is certainly desirable to examine the complete parameter space of the model to understand its phenomenology and to make plausible predictions if they can be obtained. But it will go beyond the scope of the present paper. Thus, our numerical analysis is performed using
\begin{equation}\label{eq:setparameter}
a \to 3,\ b \to 1,\ c \to 3,\ d \to -1,\ e/h \to -8/3,\ f \to 3,\ g \to 3,
\end{equation}
with such values, the matrix $M_{{\rm N},\gamma_3}^2$, eq.~(\ref{eq:matrixneutra}), is copositive and its Higgs masses eigenvalues are positive. These parameter values provide no advantage on any particular Higgs field and allowed us to have the mass of the lightest neutral Higgs to be less than 190 GeV. We can see in Figure 1 the behavior of the masses with respect to the free parameter $\omega$, and the symmetry around $\omega=0$ is evident. We found that the set of dimensionless parameter values eq.~(\ref{eq:setparameter}), gives the mass hierarchies
\begin{equation}
\begin{array}{rcl}
& M_{H_1^\pm}   & \sim  426 \ {\rm GeV} \\
400 \ {\rm GeV} < &M_{H_2^\pm} & <  670 \ {\rm GeV} \\
0 \ {\rm GeV} < &M_{H_1^0} & <  190 \ {\rm GeV} \\
200 \ {\rm GeV} < &M_{H_2^0} & <  860 \ {\rm GeV} \\
850 \ {\rm GeV} < &M_{H_3^0} & <  1000 \ {\rm GeV} \\
0 \ {\rm GeV} < &M_{H_4^0} & <  750 \ {\rm GeV} \\
850 \ {\rm GeV} < &M_{H_5^0} & <  1400 \ {\rm GeV},\\
\end{array}
\end{equation}
where $-\pi \le \omega \le \pi$. $M_{H_1^\pm}$ is constant for the set $f$, $h$ independent of $\omega$. The Higgs masses are bounded. If we calculate the average of the Higgs masses over $-\pi/2 \le \omega \le \pi/2$, we find
\begin{equation}
\begin{array}{rl}
M_{H_1^\pm}   & \sim  426 \ {\rm GeV} \\
M_{H_2^\pm} & \sim  552 \ {\rm GeV} \\
M_{H_1^0} & \sim  115 \ {\rm GeV} \\
M_{H_2^0} & \sim  567 \ {\rm GeV} \\
M_{H_3^0} & \sim  930 \ {\rm GeV} \\
M_{H_4^0} & \sim  400 \ {\rm GeV} \\
M_{H_5^0} & \sim  1167 \ {\rm GeV}.\\
\end{array}
\end{equation}

Traditionally, in the potential~(\ref{eq:potential}) the  quadratic  ($\mu_0^2$, $\mu_1^2$) and  quartic parameters ($a,b,\cdots,h$) determine the masses of the neutral and charged Higgs bosons. Otherwise, and this is the approach followed here, we can take the free parameter $\omega$ as input and determine the parameters of the potential as derived quantities. But some choices of input will lead to physically acceptable masses, $ \leq 1\, \textrm{TeV}$, and others will not.

\begin{figure}[ht] 
\centering 
\includegraphics[scale=0.7]{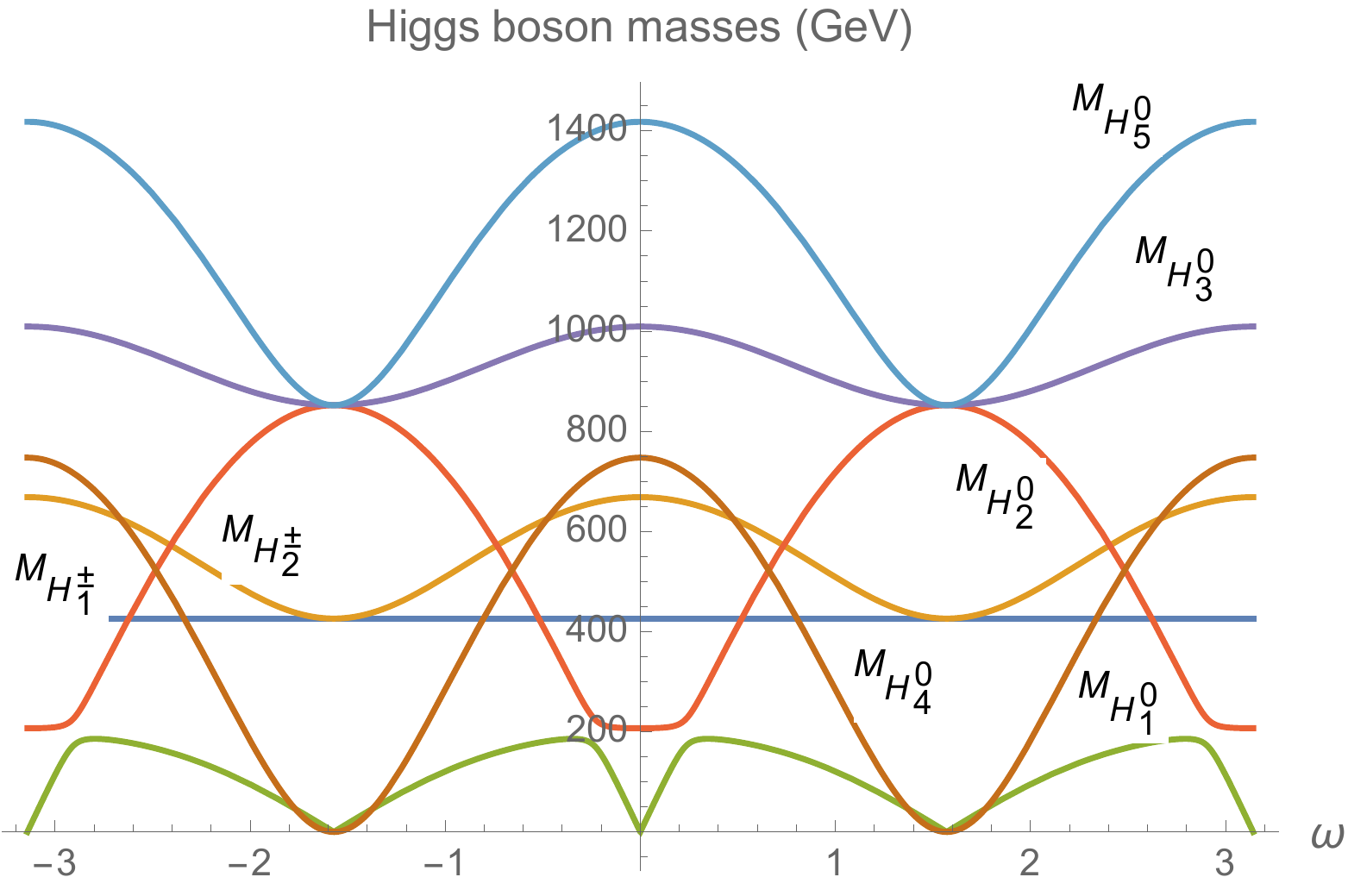}
\caption{The Higgs masses $M_{H_i^0}\, (i=1,\cdots,5)$ for $a=c=f=g=h \to 3, b=-d \to  1, -e  \to 8$, as a function of  $\omega$. In this region, 
${H_{1,4}^0}$ are candidates for a Higgs like to SM one with mass values at $125.7\pm{0.4}$ GeV, while $M_{H_{2,3,5}^0}$ will never reach this value.}  
\label{fig:masses}
\end{figure}

When analyzing the scenarios, we must consider two cases, $\omega=0, \pi/2$. We found that: $(i)$ $\omega=0$, it is the case without CPV and there are a lower Higgs masses, see Table~\ref{tab:masseshiggsII}; $(ii)$ $\omega=\pi/2$, this value constraints to the explicit CPV. Then, in Figure~\ref{fig:masses} we see that these values are meaningless. We noticed the Higgs masses only depend on $\gamma_3$; furthermore, a mass degeneration can be seen in Figure~\ref{fig:masses}, with lower masses $m_{H^0_{1}}$ and $m_{H^0_{4}}$ degenerated.

\begin{figure}[h] 
\centering 
\includegraphics[scale=.5]{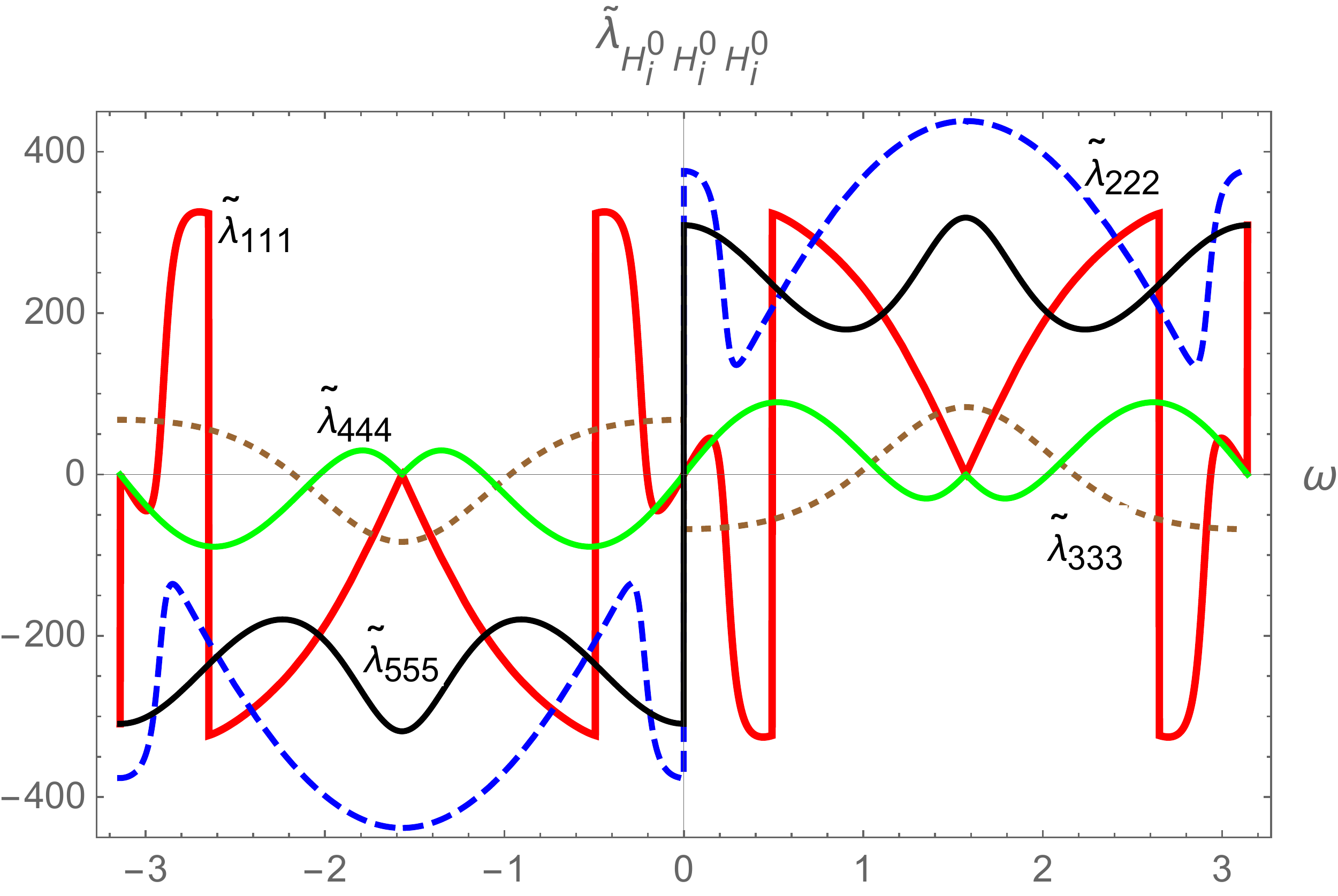}
\caption{The  trilinear Higgs self-couplings $\tilde{\lambda}_{H_i^0H_i^0H_i^0} \, (i=1,\cdots,5)$ for $a=c=f=g=h \to 3, b=-d \to  1, -e  \to 8$, in  $-\pi< \omega<\pi$. }  \label{fig:couplingh}
\end{figure}

In Figure~\ref{fig:couplingh}, the neutral Higgs self-couplings magnitudes $ \tilde{\lambda}_{H_i^0H_i^0H_i^0} $ with respect to the parameter $\omega$, and corresponding to the scenario 3 are shown, where
\begin{equation}
 \tilde{\lambda}_{H_i^0H_i^0H_i^0} \equiv \lambda_{H_i^0H_i^0H_i^0}/ \lambda_ {h ^0h^0h^0}^{SM}, \quad \lambda_ {h ^0h^0h^0}^{SM}=\displaystyle{\frac{3 M_{h^0}^{2}}{v}}.
  \label{eq:lamtilde}
 \end{equation}
%

The potential (\ref{eq:potential}) is attractive as one $S(3)$ extension of the SM that admits additional CP violation. This is an interesting possibility, since it will become possible to severely constrain or even measure it. From this potential, we can derive a function of the CP violation parameter $\gamma_3$ to the trilinear Higgs self-couplings~\cite{Barradas-Guevara:2014yoa}, which are shown in Figure~\ref{fig:couplingh}. In Figure~\ref{fig:masses}, the neutral Higgs masses with respect to the parameter $\omega$ are shown, corresponding to scenario 3, in which CP violation comes from the singlet $H_S$, for $a=c=f=g=h \to 3, b=-d \to  1, -e  \to 8$. We can observe a light Higgs with $ M_ {H_ {4}^{0}} <160 \, \textrm {GeV} $, while the others are $ M_ {H_ {1}^{0 }} <300 \, \textrm {GeV}$, and heavy Higgses with $M_ {H_ {2}^{0}} <500 \, \textrm {GeV} $, $M_ {H_ {3}^{0}} <1200 \, \textrm {GeV} $ and $M_ {H_ {5}^{0}} <960 \, \textrm {GeV} $. Further, we can see that four Higgs bosons found in a region in the parameter space reach the values of the masses of $125.7\pm {0.4}$ GeV. Each neutral Higgs acquires mass values around 125 GeV for $\omega$. Then, the computation of the self-couplings allows us to identify a Higgs like the SM one. We have to look for parameter space regions $\omega$ that simultaneously fit the Higgs mass and trilinear self-coupling for values as in the SM.

\begin{table}
\begin{tabular}{|c|c|c|c|c|c|}
\hline 
$\omega$&\multicolumn{5}{c|}{Higgs masses (GeV)}\\ \cline{2-6}
&$H^0_{1}$	&	$H^0_{2}$	&$H^0_{3}$&	$H^0_{4}$&	$H^0_{5}$ \\ \hline 
0.194	&125.258&	393.443&	1156.51&	602.474&	1313.01\\\hline
1.20	&125.472 &816.4	&993.363	&102.377&	1053.73\\\hline
1.94&	124.996&	816.764&	993.228&	101.59&	1053.21\\\hline
2.948&	125.023&	393.385&	1156.54&	602.557&	1313.04\\
\hline
\end{tabular}
\caption{Higgs masses for several $\omega$ values.}\label{tab:masseshiggsII}
\end{table}
%

%

\section{Conclusions\label{sec:conclu}}
In this work, we analyzed the SSB of SU(2) $\times$ U(1) $\rightarrow$ U(1)${}_{em}$ in $S(3)$SM with spontaneous CPV provided by the Higgs sector. In this model, we introduced three Higgs SU(2) doublets with twelve real fields. While defining the gauge symmetry spontaneous breaking in eq. ~(\ref{eq:minimcpviolation}), we found a parameter space region where the minimum of the potential defines a CPB ground state. We analyzed three possible scenarios defined in concordance with the CPV source Higgs field. Neutral and charged Higgs mass matrices were obtained for each scenario along with the eigenvalues. Thus, we found that scenario 3 contains nine massive Higgs bosons and  $W^\pm$ and $Z^0$, while scenarios 1 and 2 contain eight massive Higgs bosons and an additional Goldstone boson. Thereby, we numerically analyzed scenario 3 with nine free parameters, and we found that there are two light neutral Higgs like the SM Higgs with $m_{H^0_{1,4}}\sim 125 \, \textrm{GeV}$ for several values of $\omega$. Additionally, each value of $\omega$ gave four neutral Higgs bosons with $m > 200 \, \textrm{GeV}$, and four charged Higgs bosons with $m >400$ GeV, as the experiment points out. In this range window $M_{H_{2}^0}, \cdots, M_{H_{5}^0}$ takes smaller values to 1.4 TeV. We observed that the masses depart from zero to the maximum values. We saw that all Higgs masses are decoupled for a mass range from 110 to 140 GeV. It can be seen in Figure~\ref{fig:masses}: the Higgs masses in the range $-\pi < \omega< \pi$, where we only considered scenario 3.  Furthermore, we also computed the trilinear Higgs self-couplings $\tilde{\lambda}_{H_i^0H_i^0H_i^0}, \ i=1,\cdots,5$ as function of $\omega$. Particularly in scenario 3, we observed $H_{1,4}^0$ as possible candidates like the SM Higgs. In spite of the Higgs mass eigenvalues being positive defined, we simultaneously demand that a Higgs mass is of the order of 125 GeV and $\tilde{\lambda}_{H^0_i H^0_i H^0_i}$ of order one with the same allowed parameters point. Then, we have found that one Higgs is excluded if we consider an allowed values set, $a \to 3, b \to 1, c \to 3, d \to -1, e \to  -8, f  \to  3, g  \to 3,  h  \to 3$. For that, $2\le \tilde{\lambda}_{H_4^0H_4^0H_4^0}\le50$. At this point, we have shown the Higgs masses and trilinear self-couplings for an allowed parameters set, and shown that the Higgs mass of $H_4^0$ is sensitive to the potential parameters $f$, $g$. In this case, the trilinear Higgs self-couplings analysis confirms our hypothesis: we can have CP violation resulting from the neutral Higgs sector with a trilinear self-coupling in accordance with the SM one.

\begin{acknowledgments}
This work has been partially supported  by \textit{CONACYT-SNI (M\'exico)}. ERJ acknowledges the financial support received from \textit{PROFOCIE (M\'exico)}. The authors thankfully acknowledge the computer resources, technical expertise and support provided by the Laboratorio Nacional de Superc\'omputo del Sureste de M\'exico through the grant number O-2016/039."
\end{acknowledgments}

\appendix

\section{Scenario 1}\label{apendiceA}
The scenario 1 corresponding to $ \gamma_1\neq 0,\, \gamma_2=\gamma_3=0$,  the charged Higgs mass matrix ${\bf {M}^2_C}_{11}(\gamma_1)$ can be written as 
\be
{\bf {M}^2_C}_{11}(\gamma_1) = 
\left(
\begin{array}{ccc}
 k_4' v_2^2-\displaystyle\frac{f v_3^2}{2} & -2 d v_2 \sqrt{3 v_2^2-\gamma _1^2} & \displaystyle\frac{\left(f v_3^2-2 k_4 v_2^2\right) \sqrt{3 v_2^2-\gamma _1^2}}{2 v_3} \\
-2 d v_2 \sqrt{3 v_2^2-\gamma _1^2} & (7 d+g) v_2^2-\displaystyle\frac{f v_3^2}{2} & \displaystyle\frac{1}{2} f v_2 v_3-\displaystyle\frac{k_4 v_2^3}{v_3} \\
\displaystyle\frac{\left(f v_3^2-2 k_4 v_2^2\right) \sqrt{3 v_2^2-\gamma _1^2}}{2 v_3}& \displaystyle\frac{1}{2} f v_2 v_3-\displaystyle\frac{k_4 v_2^3}{v_3} & \displaystyle\frac{4 k_4 v_2^4}{v_3^2}-2 f v_2^2 \\
\end{array}
\right)  , \label{eq:c11g1}
\ee
where `` $\times$ '' denote the symmetric element, and $k_4 = d+g$, $k_4' = 3d+g$. We also obtained
\be
{\bf {M}^2_C}_{12}(\gamma_1) =
\left(
\begin{array}{ccc}
 0 & 2 d v_2 \gamma _1 & -\displaystyle\frac{1}{2} \left(f-\displaystyle\frac{2 k_4 v_2^2}{v_3^2}\right) v_3 \gamma _1 \\
 -2 d v_2 \gamma _1 & 0 & 0 \\
 \displaystyle\frac{1}{2} \left(f-\displaystyle\frac{2 k_4 v_2^2}{v_3^2}\right) v_3 \gamma _1 & 0 & 0 \\
\end{array} \label{eq:c12g1}
\right)  .
\ee
Using  (\ref{eq:c11g1}) and (\ref{eq:c12g1}) in (\ref{eq:matrizcargada}) we constructed the charged Higgs mass matrix. Diagonalizing this mass matrix, we obtained the charged Higgs masses:
\be
\left\{0,v_2^2 (d-2 f+g)+\displaystyle\frac{4 v_2^4 k_4}{v_3^2}-\displaystyle\frac{f v_3^2}{2},v_2^2 (9 d+g)-\displaystyle\frac{f v_3^2}{2}\right\},
\ee
then
\be
\begin{array}{rl}
M_{C1}^2 &= v_2^2 (d-2 f+g)+\displaystyle\frac{4 v_2^4 k_4}{v_3^2}-\displaystyle\frac{f v_3^2}{2} ,\\
M_{C2}^2 &= v_2^2 (9 d+g)-\displaystyle\frac{f v_3^2}{2} .
\end{array}
\ee
We have obtained four physical states of charged Higgs bosons and as we can see these masses do not dependent on $\gamma_1$ term. We have gotten two null eigenvalues to give mass to the charged vector bosons $W^\pm$. Thus,  the neutral scalar Higgs mass matrix ${\bf M}_{S}^2$, eq.~(\ref{eq:matrizneutra}), is given by
\be
{\bf {M}^2_N}_{11}(\gamma_1) = 
\left(
\begin{array}{ccc}
 2 k_1 \left(3 v_2^2-\gamma
   _1^2\right) & 2 k_{3}'
   v_2 \sqrt{3 v_2^2-\gamma
   _1^2} & \displaystyle\frac{\left(k_5
   v_3^2-4 k_4 v_2^2\right)
   \sqrt{3 v_2^2-\gamma
   _1^2}}{v_3} \\
2 k_{3}'
   v_2 \sqrt{3 v_2^2-\gamma
   _1^2} & 2 (c+6
   d+7 g) v_2^2-2 k_4 \gamma
   _1^2 & \displaystyle\frac{v_2 \left(-4
   k_4 v_2^2+k_5 v_3^2+2
   k_4 \gamma
   _1^2\right)}{v_3} \\
\displaystyle\frac{\left(k_5
   v_3^2-4 k_4 v_2^2\right)
   \sqrt{3 v_2^2-\gamma
   _1^2}}{v_3}&
\displaystyle\frac{v_2 \left(-4
   k_4 v_2^2+k_5 v_3^2+2
   k_4 \gamma
   _1^2\right)}{v_3}  &
   \displaystyle\frac{2 \left(a v_3^4+k_4
   v_2^2 \left(4 v_2^2-\gamma
   _1^2\right)\right)}{v_3^2}
   \\
\end{array}
\right), \label{eq:n11g1}
\ee
\be
{\bf {M}^2_N}_{12}(\gamma_1) = 
\left(
\begin{array}{ccc}
 2 k_1 \gamma _1 \sqrt{3
   v_2^2-\gamma _1^2} & 0 & 0
   \\
 2 k_{3}' v_2 \gamma _1 &
   2 k_4 \gamma _1 \sqrt{3
   v_2^2-\gamma _1^2} &
   -\displaystyle\frac{2 k_4 v_2 \gamma
   _1 \sqrt{3 v_2^2-\gamma
   _1^2}}{v_3} \\
 \displaystyle\frac{\left(k_5 v_3^2-4
   k_4 v_2^2\right) \gamma
   _1}{v_3} & -\displaystyle\frac{2 k_4
   v_2 \gamma _1 \sqrt{3
   v_2^2-\gamma _1^2}}{v_3} &
   \displaystyle\frac{2 k_4 v_2^2 \gamma
   _1 \sqrt{3 v_2^2-\gamma
   _1^2}}{v_3^2} \\
\end{array} 
\right), \label{eq:n12g1}
\ee
\be
{\bf {M}^2_N}_{22}(\gamma_1) = 
\left(
\begin{array}{ccc}
 2 k_1 \gamma _1^2 & 0 & 0
   \\
 0 & 2 k_4 \gamma _1^2 &
   -\displaystyle\frac{2 k_4 v_2 \gamma
   _1^2}{v_3} \\
 0 &-\displaystyle\frac{2 k_4 v_2 \gamma
   _1^2}{v_3}& \displaystyle\frac{2 k_4
   v_2^2 \gamma _1^2}{v_3^2}
   \\
\end{array}
\right) , \label{eq:n22g1}
\ee
where $k_{3}' = c-3d-2g$. We diagonalized this matrix (\ref{eq:matrizneutra}) using eqs.~(\ref{eq:n11g1}), (\ref{eq:n12g1}) and (\ref{eq:n22g1}).  We found two zero eigenstates and four nonzero mass values. We can analytically express  just  two of them, which are given by
\be\label{eq:masasg1}
\begin{array} {rl}
M_{H_1^0}^2(\gamma_1) &=
\displaystyle\frac{1}{2}
   \left({\mathcal{M}}^2_a+{\mathcal{M}}^2_c-\sqrt{({\mathcal{M}}^2_a-{\mathcal{M}}^2_c)^2+4
   {\mathcal{M}}_b^4}\right), \\[.4cm]
M_{H_2^0}^2(\gamma_1) &=   
   \displaystyle\frac{1}{2}
   \left({\mathcal{M}}^2_a+{\mathcal{M}}^2_c+\sqrt{({\mathcal{M}}^2_a-{\mathcal{M}}^2_c)^2+4
   {\mathcal{M}}_b^4}\right),
   \end{array}
\ee
where
\be
\begin{array}{rl}
{\mathcal{M}}^2_a &=
v_2 \left(v_2 (c+6 d+7
   g)-\sqrt{v_2^2 (c+6 d+7
   g)^2-4 \gamma _1^2 k_4
   (c+3 d+4 g)}\right), \\[.5cm]
{\mathcal{M}}^2_b &= \displaystyle
\displaystyle\frac{v_2}{2 v_3} \left[v_3^2
   k_5-4 v_2^2
   k_4\right. \\[.5cm]
   & \left. -\sqrt{8
   \gamma _1^2 k_4
   \left(v_3^2 k_5+2 v_2^2
   k_4\right)+\left(v_3^2
   k_5-4 v_2^2
   k_4\right){}^2}\right], \\[.5cm]
{\mathcal{M}}^2_c &= \displaystyle
\displaystyle\frac{1}{v_3^2}\left[a v_3^4+4
   v_2^4 k_4-\sqrt{\left(a v_3^4+4
   v_2^4 k_4\right){}^2-4
   \gamma _1^2 v_2^2 k_4
   \left(a v_3^4+v_2^4
   k_4\right)}\right]  .
   \end{array}
   \ee
$M_{H_{3,4}^0}^2(\gamma_1)$ have extensive expressions. All the neutral Higgs masses depend on the parameter $\gamma_1$.

By expressing the vev's of the Higgs fields as $v_i=v\cos\omega_i$ and the relationship  
\be
\begin{array}{rl}
v^2= v_1^2+v_2^2+v_3^2+\gamma_1^2.
\end{array}
\ee
In the CPB minimum for this scenario $v_1^2 = 3v_2^2 - \gamma_1^2$, then $v^2 = 4v_2^2 + v_3^2$. The masses ${M}^2_{H_i^0}$ can be parametrized with just one parameter  $\omega$.This scenario is interesting, but it has got four Goldstone bosons.

\section{Scenario 2}\label{apendiceB}
The scenario 2 corresponding to $\gamma_2 \neq 0$ and $\gamma_1=\gamma_3=0$, the charged Higgs mass matrix eq.~(\ref{eq:matrizcargada}) is written with ${\bf {M}^2_C}_{11}(\gamma_2)$ and ${\bf {M}^2_C}_{11}(\gamma_2)$, which are expressed as
\be
{\bf M_C}^2_{11}(\gamma_2) =
\left(
\begin{array}{ccc}
 k_4' v_2^2-\displaystyle\frac{f
   v_3^2}{2}+2 d \gamma _2^2 & -2
   d v_2 \sqrt{3 v_2^2+\gamma
   _2^2} & \displaystyle\frac{\left(f v_3^2-2
   k_4 v_2^2\right) \sqrt{3
   v_2^2+\gamma _2^2}}{2 v_3} \\
 -2
   d v_2 \sqrt{3 v_2^2+\gamma
   _2^2} & (7 d+g) v_2^2-\displaystyle\frac{f
   v_3^2}{2}+2 d \gamma _2^2 &
   \displaystyle\frac{1}{2} f v_2
   v_3-\displaystyle\frac{k_4 v_2^3}{v_3} \\
\displaystyle\frac{\left(f v_3^2-2
   k_4 v_2^2\right) \sqrt{3
   v_2^2+\gamma _2^2}}{2 v_3} &
 \displaystyle\frac{1}{2} f v_2
   v_3-\displaystyle\frac{k_4 v_2^3}{v_3} &
   \displaystyle\frac{\left(2 k_4 v_2^2-f
   v_3^2\right) \left(2
   v_2^2+\gamma
   _2^2\right)}{v_3^2}
\end{array}
\right),
\ee
\be
{\bf {M}_C}_{12}^2(\gamma_2) =
\left(
\begin{array}{ccc}
 0 & -2 d \gamma _2 \sqrt{3
   v_2^2+\gamma _2^2} & 0 \\
 2 d \gamma _2 \sqrt{3
   v_2^2+\gamma _2^2} & 0 &
   -\displaystyle\frac{1}{2} \left(f-\displaystyle\frac{2
   k_4 v_2^2}{v_3^2}\right) v_3
   \gamma _2 \\
 0 & \displaystyle\frac{1}{2} \left(f-\displaystyle\frac{2
   k_4 v_2^2}{v_3^2}\right) v_3
   \gamma _2 & 0
\end{array}
\right)  .
\ee

The corresponding eigenvalues for this matrix are
\be
\left\{0,v_2^2 (9 d+g)+4 d \gamma
   _2^2-\displaystyle\frac{f
   v_3^2}{2},\displaystyle\frac{\left(4
   v_2^2+v_3^2+2 \gamma
   _2^2\right) \left(2 v_2^2
   k_4-f v_3^2\right)}{2
   v_3^2}\right\} ,
 \ee
they depend on parameter $\gamma_2$ contrary to scenario 1, where there was no explicit dependence on the CP violation parameter. For the neutral Higgs mass matrix, eq.~(\ref{eq:matrizneutra}), we have    
\be
{\bf {M_N}}^2_{11}(\gamma_2) =
\left(
\begin{array}{ccc}
 2 k_1 \left(3 v_2^2+\gamma
   _2^2\right) & 2 k_{3}'
   v_2 \sqrt{3 v_2^2+\gamma _2^2}
   & \displaystyle\frac{\left(k_5 v_3^2-4
   k_4 v_2^2\right) \sqrt{3
   v_2^2+\gamma _2^2}}{v_3} \\
2 k_{3}'
   v_2 \sqrt{3 v_2^2+\gamma _2^2} & 2
   \left((c+6 d+7 g) v_2^2+k_4
   \gamma _2^2\right) & \displaystyle\frac{v_2
   \left(k_5 v_3^2-2 k_4
   \left(2 v_2^2+\gamma
   _2^2\right)\right)}{v_3} \\
 \displaystyle\frac{\left(k_5 v_3^2-4
   k_4 v_2^2\right) \sqrt{3
   v_2^2+\gamma _2^2}}{v_3} &
\displaystyle\frac{v_2
   \left(k_5 v_3^2-2 k_4
   \left(2 v_2^2+\gamma
   _2^2\right)\right)}{v_3} &
   \displaystyle\frac{2 \left(a v_3^4+k_4
   v_2^2 \left(4 v_2^2+\gamma
   _2^2\right)\right)}{v_3^2}
\end{array}
\right),
\ee
\be
{\bf {M_N}}^2_{12}(\gamma_2) =
\left(
\begin{array}{ccc}
 0 & 2 k_3 \gamma _2
   \sqrt{3 v_2^2+\gamma _2^2} &
   -\displaystyle\frac{2 k_4 v_2 \gamma _2
   \sqrt{3 v_2^2+\gamma
   _2^2}}{v_3} \\
 2 k_4 \gamma _2 \sqrt{3
   v_2^2+\gamma _2^2} & 2 (c+d+2 g) v_2 \gamma _2 & \displaystyle\frac{4
   k_4 v_2^2 \gamma _2}{v_3} \\
 -\displaystyle\frac{2 k_4 v_2 \gamma _2
   \sqrt{3 v_2^2+\gamma
   _2^2}}{v_3}  & k_5 v_3 \gamma
   _2 & \displaystyle\frac{2 k_4 v_2^3
   \gamma _2}{v_3^2}
\end{array}
\right),
\ee
\be
{\bf {M_N}}^2_{22}(\gamma_2) =
\left(
\begin{array}{ccc}
 2 k_4 \gamma _2^2 & 0 & 0 \\
 0 & 2 k_1 \gamma _2^2 &
   \displaystyle\frac{2 k_4 v_2 \gamma
   _2^2}{v_3} \\
 0 & \displaystyle\frac{2 k_4 v_2 \gamma
   _2^2}{v_3} & \displaystyle\frac{2 k_4
   v_2^2 \gamma _2^2}{v_3^2}
\end{array}
\right)  .
\ee

From here we obtained  two zero eigenvalues and four different to zero, all of them dependent on $\gamma_2$. Again, compared with the SM this scenario has an additional Higgs with zero mass. 

\section{Scenario 3}\label{apendiceC}
The scenario 3 corresponds to $\gamma_3 \neq 0$ and $\gamma_1=\gamma_2=0$,
the mass sub-matrices for charged Higgs bosons in eq.~(\ref{eq:matrizcargada}) are given by
\be
{\bf {M_C}}^2_{11}(\gamma_3) = 
\left(
\begin{array}{ccc}
 -2 g v_2^2+\displaystyle\frac{e
   \left(3 v_3^2+\gamma
   _3^2\right) v_2}{2v_3}+\displaystyle\frac{f 
   \left(v_3^2+\gamma
   _3^2\right)}{2} &
   \sqrt{3} v_2 \left(2 g
   v_2+e v_3\right) &
   \displaystyle\frac{1}{2} \sqrt{3} v_2
   \left(e v_2+f v_3\right) \\
 \sqrt{3} v_2 \left(2 g
   v_2+e v_3\right) & -6 g
    v_2^2+\displaystyle\frac{e \left(7
   v_3^2+\gamma _3^2\right)
   v_2}{2v_3}+\displaystyle\frac{f 
   \left(v_3^2+\gamma
   _3^2\right)}{2} &
   \displaystyle\frac{1}{2} v_2 \left(e
   v_2+f v_3\right) \\
\displaystyle\frac{1}{2} \sqrt{3} v_2
   \left(e v_2+f v_3\right) &
\displaystyle\frac{1}{2} v_2 \left(e
   v_2+f v_3\right) & -\displaystyle\frac{2
   v_2^2 \left(e v_2+f
   v_3\right)}{v_3} \\
\end{array}
\right), \label{eq:mc11g3}
\ee
\be
{\bf {M_C}}^2_{12}(\gamma_3) =
\left(
\begin{array}{ccc}
 0 & 0 & \displaystyle\frac{1}{2} \sqrt{3}
   v_2 \left(f+\displaystyle\frac{e
   v_2}{v_3}\right) \gamma _3
   \\
 0 & 0 & \displaystyle\frac{1}{2} v_2
   \left(f+\displaystyle\frac{e
   v_2}{v_3}\right) \gamma _3
   \\
 -\displaystyle\frac{1}{2} \sqrt{3} v_2
   \left(f+\displaystyle\frac{e
   v_2}{v_3}\right) \gamma _3
   & -\displaystyle\frac{1}{2} v_2
   \left(f+\displaystyle\frac{e
   v_2}{v_3}\right) \gamma _3
   & 0 \\
\end{array}
\right). \label{eq:mc21g3}
\ee

Now, we substituted (\ref{eq:mc11g3}) and (\ref{eq:mc21g3}) in (\ref{eq:matrizcargada}), and diagonalized the resulting matrix. The eigenvalues are 
\be
\left\{0,-\displaystyle\frac{\left(\gamma
   _3^2+4 v_2^2+v_3^2\right)
   \left(e v_2+f v_3\right)}{2
   v_3},-\displaystyle\frac{e v_2
   \left(\gamma _3^2+9
   v_3^2\right)+f v_3
   \left(\gamma
   _3^2+v_3^2\right)+16 g v_3
   v_2^2}{2 v_3}\right\}.
 \ee
The  neutral Higgs sub-matrices are given by
\be
{\bf {M_N}}^2_{11}(\gamma_3) =
\left(
\begin{array}{ccc}
 6 k_1 v_2^2 & \sqrt{3} v_2
   \left(2 k_1 v_2+3 e
   v_3\right) & \sqrt{3} v_2
   \left(2 e v_2+k_5
   v_3\right) \\
\sqrt{3} v_2
   \left(2 k_1 v_2+3 e
   v_3\right) & 2 v_2
   \left(k_1 v_2-3 e
   v_3\right) & v_2 \left(2 e
   v_2+k_5 v_3\right) \\
\sqrt{3} v_2
   \left(2 e v_2+k_5
   v_3\right) & v_2 \left(2 e
   v_2+k_5 v_3\right) & 2 a
   v_3^2-\displaystyle\frac{4 e v_2^3}{v_3}
   \\
\end{array}
\right), \label{eq:mn11g3}
\ee
\be
{\bf {M_N}}^2_{12}(\gamma_3) =
\left(
\begin{array}{ccc}
 0 & \sqrt{3} e v_2 \gamma _3
   & \displaystyle\frac{\sqrt{3} v_2
   \left(e v_2+k_5
   v_3\right) \gamma _3}{v_3}
   \\
 \sqrt{3} e v_2 \gamma _3 & -2
   e v_2 \gamma _3 & \displaystyle\frac{v_2
   \left(e v_2+k_5
   v_3\right) \gamma _3}{v_3}
   \\
 -\displaystyle\frac{\sqrt{3} e v_2^2
   \gamma _3}{v_3} & -\displaystyle\frac{e
   v_2^2 \gamma _3}{v_3} & 2 a
   v_3 \gamma _3 \\
\end{array}
\right), \label{eq:mn12g3}
\ee
\be
{\bf {M_N}}^2_{22}(\gamma_3) =
\left(
\begin{array}{ccc}
 -\displaystyle\frac{v_2 \left(2 k_4 v_2
   v_3+e \left(v_3^2+\gamma
   _3^2\right)\right)}{v_3} &
   \sqrt{3} v_2 \left(2 k_4
   v_2+e v_3\right) & 0 \\
 \sqrt{3} v_2 \left(2 k_4
   v_2+e v_3\right) &
   -\displaystyle\frac{v_2 \left(6 k_4
   v_2 v_3+e \left(3
   v_3^2+\gamma
   _3^2\right)\right)}{v_3} &
   0 \\
 0 & 0 & 2 a \gamma _3^2 \\
\end{array}
\right). \label{eq:mn22g3}
\ee

We computed the neutral matrix (\ref{eq:matrizneutra}) with (\ref{eq:mn11g3}), (\ref{eq:mn12g3}) and (\ref{eq:mn22g3}). Diagonalizing the resulting matrix, the eigenvalues are: one zero and five non zero, there are only three Goldstone bosons. When analyzing the  Higgs masses for these three scenarios,  we see again that in scenario 3 the mass spectrum of Higgs bosons is obtained analogous to the normal minimum, where CP is conserved. For this, we have four electrically charged Higgs bosons, with degenerated masses, two by two, five neutral bosons, and three massless bosons, which are given mass to vector bosons. The eigenvalues are shown in Figures~\ref{fig:masses}.

%

\end{document}